*Original Article*

# Hybrid Key Authentication Scheme for Privacy over Adhoc Communication

B. Murugeshwari[1], R. Amirthavalli[2], C. Bharathi Sri[3], S. Neelavathy Pari[4]

[1,2,3]Velammal Engineering College, Chennai, India
[4]Department of Computer Technology, MIT Campus, Anna University, Chennai, India

[1]Corresponding Author : niyansree@gmail.com

***Abstract*** *- Since communication signals are publicly exposed while they transmit across space, Ad Hoc Networks (MANETs) are where secured communication is most crucial. Unfortunately, these systems are more open to intrusions that range from passive listening to aggressive spying. A Hybrid Team centric Re-Key Control Framework (HT-RCF) suggests that this research examines private group communication in Adhoc environments. Each group selects a Group Manager to oversee the group's members choose the group manager, and the suggested HT-RCF uses the Improved Hybrid Power-Aware Decentralized (I-HPAD) mechanism. The Key Distribution Center (KDC) generates the keys and distributes them to the group managers (GMs) using the base algorithm Rivest Shamir Adleman (RSA). The key agreement technique is investigated for safe user-user communication. Threats that aim to exploit a node are recognized and stopped using regular transmissions. The rekeying procedure is started every time a node enters and exits the network. The research findings demonstrate that the suggested approach outperforms the currently used Cluster-based Group Key Management in terms of power use, privacy level, storage use, and processing time.*

***Keywords*** *- Hybrid Team centric Re-Key Control Framework (HT-RCF), Key management, RSA, MANET.*

## 1. Introduction

MANET is a transient, multi-hop independent system of numerous mobile nodes that lack a centralized control structure or fixed architecture. MANETs are frequently utilized in emergencies, mobile conferencing, sensor networks, and other applications [1-4]. MANET is a novel class of wireless communication network with features including dynamic topology, lack of a central control point, network fragility, constrained communication capacity, and host energy. The security issues faced by mobile ad hoc networks are more significant than those faced by conventional fixed networks. The topology of a MANET varies dynamically, there is no central command and control point, and the network is open to intrusion. An essential factor in ensuring the privacy and security of information is "key" management on nodes in the network.

The fundamental problem with the security of mobile ad hoc networks, which has drawn a lot of attention [5-8], is key management. In ad hoc communication, some research suggested decentralized key distribution techniques in which an authorized center generates each networking node's sub-key. However, the single-node trusted center key management strategy will likely result in the single point of failure issues in mobile ad hoc networks, meaning that if one node is attacked, the network could be vulnerable [22-25]. If there is widespread trust in the system, the system can continue to operate well and be accessible even if certain nodes malfunction or leak.

Overall current mobile networks, One of the more important and distinctive applications is called MANET (Mobile Ad hoc Network), allowing for sharing files, using sensing devices, and vehicle-to-vehicle transmission. Key control is crucial to safeguarding connectivity in large, unsecured networks [3]. Since smartphone apps are becoming more widespread and commonplace, safe data transfer is the primary necessity from a real-time technological standpoint. To provide telehealth for essential technologies such as secure SMS short messaging transfer for health-related issues, secure business agricultural monitoring, secure business heart illnesses, secure army monitoring and situations, and secure communications, for instance, it is necessary to secure identity management and distribution surroundings and security, the way of the system. Key management and wireless connectivity for mobile devices [9-11].

Clustering is a way to get around the problems of a central controller and a lack of resources in ad-hoc networks. Cluster network approaches offer improved scalability and dependability compared to simple techniques [12]. However, efficient power utilization is a significant concern in wireless ad hoc systems. A Hybrid Team Centric Re-Key Control Framework (HT-RCF) advises that this research look at private group communication in Adhoc situations as a solution to this issue [26]. The suggested HT-RCF uses the Improved Hybrid Power-Aware Decentralized (I-HPAD) method, in which each group chooses a Group Manager responsible for overseeing the group members.

## 2. Related Work

The network's routing protocol governs and controls how data messages are forwarded between nodes. Additionally, clustering is a crucial technique in WSNs to localize computation and reduce transmission overhead (Wang et al.,2010) [11]. Clustering is the most common process of combining and selecting a set of group managers or group leaders from each sensor network is the primary activity in sensor network grouping (Klaoudatou et al.,2011) [4]. Organization among nodes and inter-node communication are the responsibilities of the group controller nodes. Numerous routing and key management techniques have already been developed. Each group has a group management node, while the cluster normally has a cluster head node. The following key differences exist between combining and clustering: While grouping often concentrates on a particular area, clustering is a ubiquitous concept. A group could be a combination of numerous clusters or a portion of the group.

The most significant area of study in wireless communication is security. The trustworthy transfer of the group key between active nodes in a cluster is the backbone of secure group interaction. It also covers the exchange of communications among nodes that are permitted to transmit and receive messages for the group. Key management is a significant issue with secure group interaction. Data communication between group users is encrypted using an asymmetrical group key only known by the key server and group nodes. The group key control system regulates group key accessibility. When a node enters or leaves the system, it rekeys it and provides the private key to the authorized new users. (Wen et al., 2012) [12] Specifically, a group key distribution architecture can implement reverse access or forward authentication protocols (Alzaid et al., 2010) [1]. There is a chance that the new node cannot decrypt earlier messages if the group changes the subgroup key after node mobility joins; this is known as backward network access. Forward access management prevents a departing node from accessing future communications if the grouping rekeys after a current iteration leaves.

Rani et al. [15] suggested a protected encryption system and secured key distribution for group interaction. Every node in the cluster needs to keep a key pair and a shared key for interaction. The B-Tree architecture was used for rekeying purposes in that research during node mobility to create a new group key. Concerning the distribution of group keys, such a tactic reduced communication costs. The Symmetric Key Encryption method given was simple and efficient regarding communication costs. This GKM with classification methods was represented using NS2.

Making a proposed Identity-Based Key Agreements System by Yang et al. in 2013 [13]. (IBKAS). Elliptic Curve Diffie Hellman and identity-based cryptography were the foundation for this scheme (ECDH). By encryption of the key establishment factors, these technique guards against node takeover and man-in-the-middle – attack. A key distribution system for sensor systems based on the Diffie Hellman group by Lin et al. (2010) [7]. Any node couple can consistently give one session key thanks to this scheme. Finally, a cluster-based method to group access control for WSN by Yuan et al. [14]. A group key was produced by connecting the cluster center and individual cluster nodes. Cluster members were responsible for reassembling and transferring the group key. The secure communication method has the benefit of preventing external nodes from receiving messages. In addition, the group processing center gathers comparable data to avoid transmission overhead. Recent studies have shown that the group key can filter the incorrect data present in WSN [6].

Mohindra and Gandhi [16] suggested an energy-efficient technique for ad-hoc networks for cooperative packet forwarding. The connection availability measurements and the CH node's handling capacity were selected. CH receives data from the servers, transmitting it at node connection speeds to the cluster members. The data offloading approach in networking controls the data traffic. The CH reconnection technique integrates load balancing in a network. The given clustering technique significantly reduced the data flow and power usage, and then through reliable routing, the bandwidth rate was increased.

Anupama and Sathyanarayana developed a safe and effective "key" management method in MANET [17]. First, mobile input nodes were selected using a soft computing method. Then, the nodes have grouped using the FCM (Fuzzy C-means grouping) algorithm. Next, the clustering nodes were optimized for choosing the correct number of nodes at the right time for communicating. The EBFO (Extended Bacterial Foraging Optimization) technique can do the optimisation. Finally, the researchers used the Elliptic Curve Diffie-Hellman, or ECDH, authentication method. That "key" agreement method distributes symmetrical keys among the participants to maintain anonymity at a lower cost. In addition, the networking overhead was kept to a minimum by employing ECDH.

AMGKM (area-based multiple GKM) was a novel GKM (Group key maintenance) schema that Zhong et al. [18] created for multiple broadcast groups on the internet of individuals. The AMGKM was implemented and also allowed people to navigate across wireless connections over the Web of people. At the same time, clients subscribed to various multicast services with decreased transmission overheads in wireless mobile service scenarios, such as VANETs. Additionally, using that schema, the complexities of the signaling burden might be effectively reduced, and both reverse and forward security and supporting members' changes could be satisfied. The authors' proposed technique takes advantage of the MKE (master key encryption) technique to provide intense protection while efficiently reducing the rekeying overhead cost.

Rao et al. [44] presented a power-aware PSO-based CH selection method using an optimum particle description and heuristic information. The authors calculated the distance inside the blocks, the distance, and the remaining energy of the terminals for the algorithm's power generation. Finally, the authors use the weighting function to build a group. The algorithm was extensively evaluated under numerous WSN scenarios and conditions and contrasted with empirical results and other current algorithms, including Leach, LE-C, PSO-C, LDC, and ELeach. It was discovered that the provided technique outperformed the existing approach very well. As a result, the ground station receives overall network longevity, power usage, and datagram count. Unfortunately, the developers of the provided approach did not consider any routing algorithms.

The effectiveness of the suggested EIMKM: Enhanced Multiple Key Management Strategy, the ID-based Multiple Secret Key Management Framework (IMKM), the Centrally controlled Flat Table Key Management (CFKM), the Elgamal Key Exchanging Methodology (ElGamal), the Scalable Multihop Key Dispersion (SMKD), and the existing Elgamal Key Managerial (ElGamal) procedures has been tested by Mohandas and Krishnamoorthi [20]. According to the analytical outcomes, the provided EIMKM Key Management System outperformed existing IMKM and other existing strategies in terms of increased movement, key update duration, key sharing duration, safety strength, messaging overhead, computation cost, and energy consumption.

## 3. Proposed Methodology

Key distribution techniques for one-to-all communication are necessary for constructing a system with a secret key by the subgroup network participants. Creating pair-wise keys distributed by all group mates and group managers requires the key generation for one-to-one communication supporting aggregation. All clusters also distribute the group key.

Assume that there are n nodes in the network. We use the Improved Hybrid Power-Aware Decentralized (I-HPAD) method for personal group interaction in Adhoc contexts. A group comprises n-1 groupmates and one Group Manager. Each group chooses a Group Manager by voting to monitor the group's conduct. Each cluster has a distinct, independent group ID (IDg), and each node has a unique node ID (IDn). Each member of the group has access to a subgroup key called kg. The group members use this key to access personal conversations and data sharing [27-30]. The suggested system enables data transmission protection based on a group key that distributes between team members.

Without the assistance of a centrally controlled or global key manager, the Key Distribution Centre (KDC) enables the creation and transmission of keys within subgroups of any size. It generates the group key and distributes it by a group manager using the unique keys belonging to each member, and it knows only by the group's active users in KCD. The challenges associated with users entering or leaving are simple [31-35]. First, the manager transmits the more significant set key to the entering user across the communication channel and encrypts it all with the old group key before sending it to the nodes in the old subgroup.

### 3.1. Cluster Forming and Privacy Requirements

Consider the nodes to be partially stable. Nodes can communicate and interpret information similarly. The I-HPAD method is used to select the group manager. Since it controls all groupmates, the group manager can determine which nodes have been attacked by enemy nodes. An IDn identifies each node, and each node may be a member of more than one subgroup. Each node regularly transmits the Beacon signal to the cluster to signal its existence [36-38].

**Table 1. Notation and meanings**

| S.No | Notations | Meanings |
|---|---|---|
| 1. | $ID_n$ | Independent Node Unique ID |
| 2. | $ID_g$ | Independent Group ID |
| 3. | $Gr_p$ | Group probability |
| 4. | $GM_p$ | Group Manager Probablity |
| 5. | $P_{Res}$ | Power Residuals |
| 6. | $P_{Ext}$ | Extreme Power / Maximum Power |
| 7. | $Key_{pub}$ | Public Key |
| 8. | $Key_{Gr}$ | Group Key |
| 9. | $Key_{pri}$ | Private Key |
| 10. | $Key_{sec}$ | Secret Key |

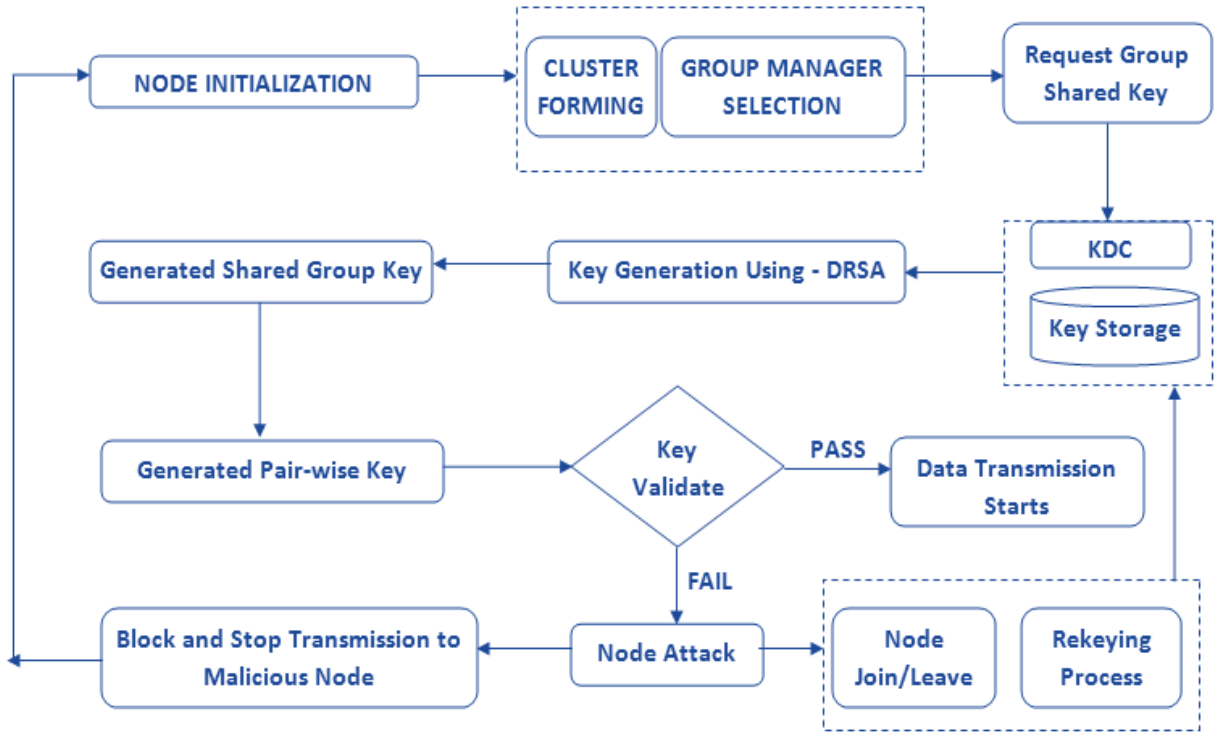

Fig. 1 Architecture of Proposed Model

The node will be labeled malicious if it fails to transmit any signals. These nodes will be placed on a delisted to enable safe data exchange. With this technique, both forward as well reverse anonymity is achieved.

### 3.2. HT-RCF with I-HPAD techniques
The multi-hop Ad hoc communication and node grouping algorithm HT-RCF with I-HPAD offers power-aware routing with a clear focus on performance. The following two factors are used to achieve the group dynamics:

- Nodes leftover power
- Cost of intra - team transmission

Since an ideal percentage cannot be established beforehand, the fraction of GM over all nodes, $Gr_p$, is predetermined. The likelihood that the node will choose GM:

$$GM_p = \frac{G_p \times P_{Res}}{P_{Ext}} \quad (1)$$

Here, $P_{Ext}$ is the highest power, and $P_{Res}$ is the calculated power of the present node. There is a minimum value under which $GM_p$ cannot be permitted to fall. The chosen thresholding has an opposite relation to $P_{Ext}$.

Each node makes several cycles before choosing the GM. Then, the node votes itself to be GM and gives a signal to its neighboring nodes if no GM is chosen.

Once GM hits 1, each node doubles its $GM_p$ value before moving on to the following iteration.

A mobile node could transmit one of two statuses to its neighbors:

$$\begin{cases} Tentative & G_p < 1 \\ Final & G_p = 1 \end{cases} \quad (2)$$

The node chooses GM indefinitely.

The following are the benefits of the I-HPAD procedure:
a) The I-HPAD clustering technique thoroughly distributes the effectiveness of two variables for GM voting.
b) While inter-group interaction involves extensive power levels for groupings, reduced power levels of clusters encourage a rise in spatial reuse. As a result, the GM distributes equally, and the burden is balanced.
c) It successfully strikes a balance between single or multi-hop interactions.

### 3.3. Key Generation and Distribution Center (KDC)
The key management system is a collection of processes and tools that facilitate the creation of a shared secret key and control the existing key relationship among nodes by, if necessary, exchanging outdated keys with new ones [39]. Only verified nodes should be able to access the network or participate in conversations, according to KDC. To prevent the new team members from being unable to view the previous data flow, the public key should be changed whenever a network node shifts. Rekeying begins when a node enters or exits the system [40].

1. Shared Group Key Generation- Device-Based RSA(DRSA)

The KDC receives the group key req from the group manager. In order to send the secured key to the appropriate group manager, KDC produces the group key req, which is made up of the independent key and the group key. The RSA algorithm is the base for producing the keys.

**DRSA - Algorithm for Key Generation**
Start
Step 1: Choose two distinct prime numbers based on the device's unique ID u and v
Step 2: Initialize the two prime
Step 3: Compute $n = uv$
Step 4: Compute $\emptyset(n) = \emptyset(u)\,\emptyset(v) = (u-1)(v-1)$
Step 5: Choose an integer k --> $1 < k < \emptyset n$
Step 6: $gcd(k, \emptyset(n)) = 1$
Step 7: Find l as $l^{-1} = k \,(mod\, \emptyset(n))$
Step 8: Publish the public encrypt key: $(k, n)$
Step 9: Keep secret private decrypt key: $(l, n)$
Step 10: Stop the process
Compute Cipher text and Plain Text

$$\begin{cases} c = m^e (mod\, n) & 0 \le m < n \\ m = c^d (mod\, n) & 0 \le m < n \end{cases} \quad (3)$$

The Group managers receive the keys after they have been produced. The keys are given out by group managers to the participants of that specific group [41]. The Diffie Hellman technique is used in the suggested systems to transfer and verify the keys. This technique verifies the authenticity of the two participating nodes.

### 3.4. Key Verification and Transmission using the Diffie-Hellman technique

The suggested system uses the Diffie-Hellman technique to transfer and verify user keys. Think about the source (sr) and recipient (re) nodes.

Algorithm for Key Verification and Transmission:
Step 1: sr selects a random key as $Key_{priM} = m \in \{1,2,..,x-1\}$.
Step 2: re-select a random key as $Key_{priN} = n \in \{1,2,..,x-1\}$.
Step 3: sr calculates the corresponding public key
$Key_{pubM} = M = m^m \,mod\, x$
Step 4: re calculates the corresponding public key
$Key_{pubN} = N = m^n \,mod\, x$
Step 5: sr calculates the common secret $Key_{MN} = N^m = (m^m)^n \,mod\, x$
Step 6: re calculates the common secret $Key_{MN} = M^n = (m^n)^m \,mod\, x$
Step 7: Use the joint key $Key_{MN}$ for encryption (with RSA)
$B = DRSA_{KeyMN}(b) \to A = DRSA^{-1}{}_{KeyMN}(b)$

The node recognizes as a trustworthy node, and data flow to the user is permitted if the key matches. If not, the rekeying procedure starts, and the node is labeled dangerous.

### 3.5. Privacy with Rekey Mechanism

A group key is initially determined and obtained from KDC. Then, the new group key is communicated to the incoming team member through the secure connectionless channel without disclosing the sender's identity and location information. When a team member departs the group, the new group key should be secretly transmitted to all surviving participants, excluding the departing node. Since the group is changeable, the key needs to be adjusted accordingly. This procedure must be accomplished with privacy.

When a node enters or departs the network, the group control system rekeys the node and transmits the group key to verify new users [42]. The reverse authentication process and forward authentication protocols are access control types with an authenticated key agreement paradigm [43]. For example, the new node might not be able to decipher earlier messages if the group changes the shared key after the new node enters; this is known as reverse network access. It is known as forwarding access control. Likewise, if the subgroup rekeys after a node departs, the departing node may not be able to obtain the subsequent messages.

The rekeying procedure depends on three basic types of information as follows:

1. Every node in the network has a unique secret key ($K_i$).
2. GM should not share and use rekeying data to estimate a beacon signal.
3. To determine the new group key, GM uses the beacon signal along with $K_i$.

A new key must be produced to avoid future interaction and preserve the previous conversation when a group member enters or quits the group. To ensure that only trustworthy members receive these rekeyed pieces of information and that banned members cannot obtain the rekeyed, the rekey operation specifies how well the new keys are distributed to nodes.

## 4. Performance Analysis

This section assesses and contrasts the performance of the suggested approach with that of the existing system. We formulate a system of 100 nodes with a hybrid team-centric rekey control scheme, and nodes are spread out over a $100 \times 100$-meter region using a homogeneous distribution channel.

The network elements are made up of several I-HPAD-based groups. Various analyses like power consumption, privacy level assessment, key accurateness, memory utilization, and time analysis are used to evaluate performance. Furthermore, according to the Hybrid Team Centric Re-Key Control Framework (HT-RCF), this study should look at private group communication in Adhoc contexts. Therefore, each group chooses a Group Manager to govern the group. The group members suggest Improved HT-RCF using Hybrid Power-Aware Decentralized (I-HPAD) technique compared with the existing protocol suggested by Danni Liu et al. [21].

### 4.1. Power Utilization

Security procedures use power via analysis and processing and absorb power during data communication. The source address and the receiver node determine the transmission power level and power usage. Therefore, the amount of time needed to send a message is inversely proportional to the concentration of the information and directly related to the transmission rate as follows:

$$T_{pow} = \Sigma S_{pow} + \Sigma R_{pow} \qquad (4)$$

$T_{Pow}$ stands for total power consumed, $R_{pow}$ for available power used during data receiving, and $S_{pow}$ for total power consumed during data transmission.

The Improved HT-RCF protocol uses the proposed Hybrid Team Centric Re-Key Control Framework to generate subgroups and choose group managers. The main restriction for choosing the group manager is residual power.

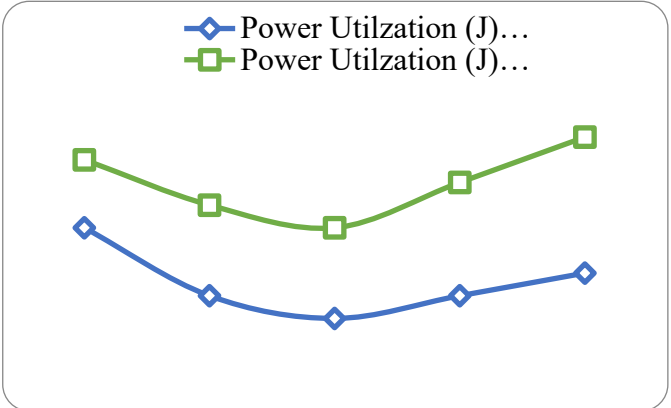

**Fig. 2 Power Utilization (J) – HT-RCF vs Danni [21]**

**Table 2. Power Utilization (J) – HT-RCF vs Danni [21]**

| Details | Power Utilzation (J) | |
|---|---|---|
| | Proposed HT-RCF | Danni Liu et al. [21] |
| Group 1 | 8 | 11 |
| Group 2 | 5 | 9 |
| Group 3 | 4 | 8 |
| Group 4 | 5 | 10 |
| Group 5 | 6 | 12 |

The above graph clearly states that the power utilized by our proposed model is high, and each group utilizes less power when compared to the existing model. It shows that every group utilizes nearly 45 to 50 % of the power.

### 4.2. Privacy Level Assessment

The proposed technique securely distributes the group key to the appropriate group members via the group manager whenever a node enters or departs. In addition, the number of repetitions is explored along with the privacy level. The practical findings demonstrate that similar to the previous Key Control Framework by Danni et al. [21], the suggested HT-RCF ensures a high degree of privacy.

**Table 3. Privacy level analysis among HT-RCF vs Danni [21]**

| No of Execution | Privacy Level (bits) | |
|---|---|---|
| | Proposed HT-RCF | Danni Liu et al. [21] |
| Round 1 | 10 | 4 |
| Round 2 | 12 | 5 |
| Round 3 | 9 | 3 |
| Round 4 | 11 | 4 |
| Round 5 | 13 | 5 |
| Round 6 | 15 | 6 |

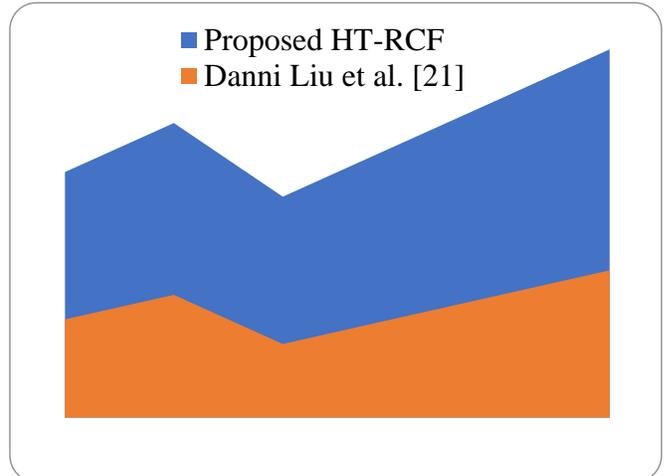

**Fig. 3 Privacy level analysis among HT-RCF vs Danni [21]**

**Table 4. Accuracy level analysis among HT-RCF vs Danni [21]**

| No of Iterations | Accuracy Level (bytes) | |
|---|---|---|
| | Proposed HT-RCF | Danni Liu et al. [21] |
| Round 1 | 51 | 38 |
| Round 2 | 56 | 26 |
| Round 3 | 47 | 24 |
| Round 4 | 85 | 37 |
| Round 5 | 46 | 15 |
| Round 6 | 78 | 24 |

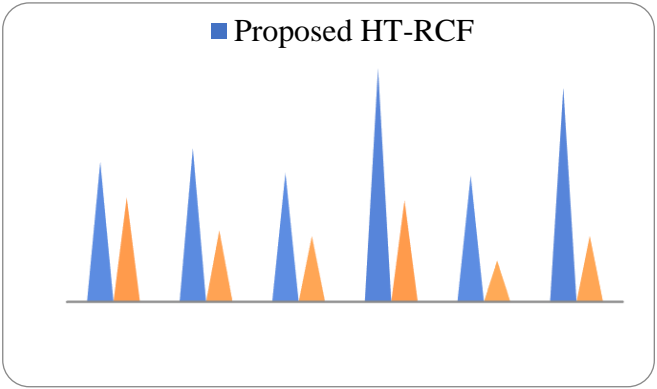

**Fig. 4 Accuracy level analysis among HT-RCF vs Danni [21]**

### 4.3. Key Accurateness

Key accuracy levels for projected HT-RCF and Danni [21] are contrasted and assessed. The present scheme used the Diffie Hellman algorithm for key exchange among nodes. The suggested system outperforms the current method regarding key accuracy, as demonstrated in Fig. 4.

### 4.4. Memory Utilization

The memory use is examined and contrasted with the previous Danni [21] approach and suggested HT-RCF in Fig. 5. It demonstrates that the suggested plan uses less RAM than the previous strategy.

**Table 5. Memory Utilization between HT-RCF vs Danni [21]**

| No of Iterations | Memory Utilization (bytes) | |
|---|---|---|
| | Proposed HT-RCF | Danni Liu et al. [21] |
| Round 1 | 11 | 54 |
| Round 2 | 14 | 66 |
| Round 3 | 21 | 83 |
| Round 4 | 14 | 75 |
| Round 5 | 20 | 79 |
| Round 6 | 23 | 78 |

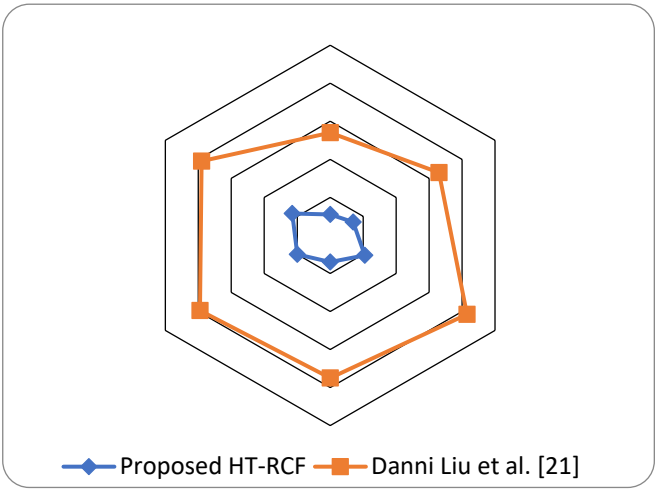

**Fig. 5 Memory Utilization between HT-RCF vs Danni [21]**

### 4.5. Time Analysis

Fig. 6 shows the typical time needed to simulate the current Danni [21] and HT-RCF. Again, the suggested model runs faster than the current model.

$$T_{Time} = \Sigma S_{Time} + \Sigma R_{Time} \qquad (4)$$

S$_{\_Time}$ is the amount of time Source takes to achieve the data transmission, R$_{\_Time}$ is the amount of time it took to receive the data, and T$_{\_Time}$ is the amount of time it took to send the information.

**Table 6. Time Consumption HT-RCF vs Danni [21]**

| No of Iterations | Time Consumption (ms) | |
|---|---|---|
| | Proposed HT-RCF | Danni Liu et al. [21] |
| Round 1 | 32 | 48 |
| Round 2 | 26 | 57 |
| Round 3 | 21 | 45 |
| Round 4 | 31 | 67 |
| Round 5 | 12 | 34 |
| Round 6 | 23 | 54 |

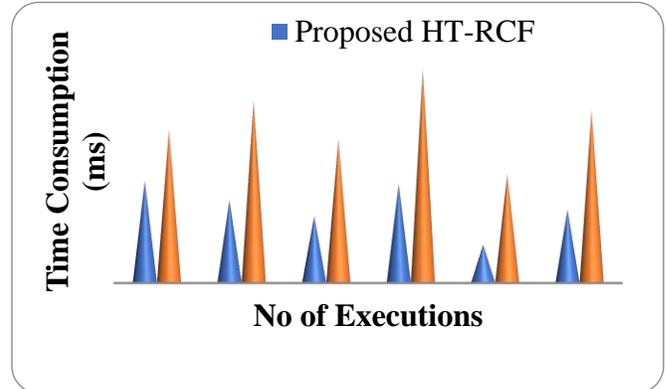

**Fig. 6 Time Consumption HT-RCF vs Danni [21]**

## 5. Conclusion

This research suggests a hybrid team-centred rekey control framework. The grouping component of this approach separates the network into smaller groups using group clusters. This strategy takes advantage of the rekeying procedure to resolve the abovementioned issue. Rekeying phase is how both forward and reverse privacy are made possible. As a result, the shareable keys are regularly updated. The periodical beacon signal sent by the team members allows for detecting and preventing the node replication assault in the sensor field. According to the experimental results, the proposed system outperforms the current model regarding power utilization, privacy level, key accuracy, memory utilization, and time consumption. In the future, Trust Centric Re-key control protocols can use the suggested technique to identify malicious attacks and efficiently increase network efficiency.